\documentclass[
    ,final            
    ,numberedheadings 
    ,cmfonts          
  ]
  {aipproc}

\layoutstyle{6x9}
\usepackage{amssymb}

\newcommand{\be}{\begin{eqnarray}}
\newcommand{\ee}{\end{eqnarray}}
\newcommand{\el}{\nonumber \hfill \\}
\newcommand{\Tr}{\mathrm{Tr}}

\begin{document}

\title{Volume Dependence of the Pion Mass from Renormalization Group Flows}


\author{B. Klein\footnote{email: b.klein@gsi.de}\ }{
  address={GSI, Planckstrasse 1, 64291 Darmstadt, Germany},
altaddress={Institute for Theoretical Physics, University of
  Heidelberg, Philosophenweg 19, 69120 Heidelberg, Germany}
}

\author{J. Braun}{
  address={Institute for Theoretical Physics, University of
  Heidelberg, Philosophenweg 19, 69120 Heidelberg, Germany}
}

\author{H.J. Pirner}{
  address={Institute for Theoretical Physics, University of
  Heidelberg, Philosophenweg 19, 69120 Heidelberg, Germany},
altaddress={Max-Planck-Insitut f\"ur Kernphysik, Saupfercheckweg 1,
  69117 Heidelberg, Germany} 
}
\begin{abstract}
We investigate finite volume effects on the pion mass and the pion
decay constant with renormalization group (RG) methods in the
framework of a phenomenological model for QCD. An understanding of
such effects is important in order 
to interpret results from lattice QCD and extrapolate reliably from finite
lattice volumes to infinite volume. 

We consider the quark-meson-model in a finite Euclidean $3+1$ dimensional
volume. In order to break chiral symmetry in the finite volume, we introduce a
small current quark mass. In the corresponding effective potential for
the meson fields, the chiral $O(4)$-symmetry is broken explicitly,
and the sigma and pion fields are treated individually.
Using the proper-time renormalization group, we derive
renormalization group flow equations in the finite volume and solve
these equations in the approximation of a constant expectation value.

We calculate the volume dependence of pion mass and pion decay
constant and compare our results with recent results from chiral perturbation
theory in finite volume.
\end{abstract}

\maketitle


\section{Introduction}
Two important motivations inspire us to study QCD in a
finite volume: a generic interest in finite volume effects, and the
separation of scales afforded by the additional length scale. Among
nonperturbative methods used in the 
study of QCD, lattice simulations have a very important place. However,
they are necessarily performed in relatively small volumes, and
a good understanding of finite volume effects is required for an
extrapolation to infinite volume.
Typically, lattice sizes are of the order of $L \lesssim 2 - 3 \;
\mathrm{fm}$, while pion masses are of the order $m_\pi \gtrsim 500 - 700
\;\mathrm{MeV}$ \cite{AliKhan:2003cu, Aoki:2002uc}.
Thus, we require tools to describe the volume dependence of
observables in these regimes. 
In this work, our objective is to describe the volume dependence of
two of the low
energy observables, the pion mass and the pion decay constant, and to
provide a model for finite volume effects across a wide range of
length scales.  

The low energy behavior of QCD is determined by spontaneous chiral
symmetry breaking \cite{Gasser:1983yg}: As a consequence, massless
Goldstone bosons associated with the broken symmetries emerge, and the
low energy limit can be described by an effective theory of these
light, weakly interacting degrees of freedom.
Such descriptions in terms of light degrees of freedom only become
even better in finite Euclidean volumes: contributions of heavier
particles are suppressed by $e^{-ML}$, where $M$ is the typical
separation of the hadronic mass scale from the Goldstone masses.  
Therefore, a second motivation to study finite volume QCD is
that it allows isolation of the low-energy behavior 
and a description in terms of the Goldstone modes. 

Different approaches based on this idea have proven very fruitful: Chiral
perturbation theory
makes predictions for the volume and temperature dependence of pion
mass, pion decay constant and chiral condensate
\cite{Gasser:1986vb, Gasser:1987ah, Gasser:1987zq}, finite volume
partition functions \cite{Leutwyler:1992yt} and random matrix theory
\cite{Shuryak:1992pi, Verbaarschot:1994qf} predict eigenvalue spectra, 
and the dependence of the chiral condensate on volume and quark mass
\cite{Verbaarschot:1995yi}.
 
The most important results in the present context are those from
chiral perturbation 
theory (chPT), which relies on an expansion in terms of the three-momentum
$|\vec{p}|$ and the pion mass $m_\pi$, which are small compared
to the chiral symmetry breaking scale $4 \pi f_\pi$. Consequently, a
finite volume places constraints on the expansion, since the
smallest momentum is determined by the volume, $p_{\mathrm{min}} \sim
\frac{2 \pi}{L}$, and requires different expansion schemes, depending
on the relative size of $L$ and $1/m_\pi$. 

A very useful result obtained by L\"uscher \cite{Luscher:1985dn}
relates the leading corrections to the pion mass in
finite Euclidean volume to the $\pi\pi$-scattering amplitude in infinite
volume. The relative shift $R[m_\pi(L)]$ of the pion mass $m_\pi(L)$ in finite
volume compared to its value $m_\pi(\infty)$ in infinite volume is
according to this result given by 
\be
R[m_{\pi}(L)] &=& \frac{m_\pi(L) - m_\pi(\infty)}{m_\pi(\infty)}\el 
               &=& -\frac{3}{16 \pi^2}\frac{1}{m_\pi} \frac{1}{m_\pi L} 
                   \int_{-\infty}^{\infty} dy\, 
F(i y) e^{-\sqrt{m_\pi^2+y^2} L}+{\mathcal O}(e^{-\bar m L}).
\label{eq:Luscher}
\ee
$F(s)$ is the forward $\pi\pi$-scattering amplitude as a function of
the energy variable $s$, continued to complex values, and the
sub-leading corrections drop at least as ${\mathcal O}(e^{-\bar m L})$ where
$\bar m \ge \sqrt{3/2}\, m_\pi$. New results have been obtained by
using a calculation of the $\pi\pi$-scattering amplitude in chPT
to three loops ({\it nnlo}) as input for L\"uscher's formula
\cite{Colangelo:2003hf, Colangelo:2004xr}. The shift above the leading one-loop
result from this approach
can then be used to supplement the full one-loop calculation
\cite{Gasser:1987ah} for the mass shift in chPT. 

However, this explanation relies purely on the ``squeezing'' of a
``pion cloud'' \cite{AliKhan:2003cu}, whereas - albeit smaller -
shifts of meson and hadron masses
also appear in quenched lattice calculations, show a power law
behavior for small $L$ rather than the behavior expected from chPT
\cite{Aoki:1993gi, Fukugita:1992jj}, 
and are generally underestimated \cite{Aoki:2002uc}.  
This suggests that in addition to the pion effects, quark effects at
higher momentum scales also contribute directly to the finite volume
mass shifts.    

In the present investigation, we thus use a phenomenological low-energy
model which is suited to a description of dynamical chiral symmetry
breaking and which incorporates heavy constituent quarks. 
As a drawback, this model is not a gauge theory and the
constituent quarks 
are not confined, but decouple from the dynamics at low
momenta only because of their large masses.
Since spontaneous symmetry breaking does not occur in a finite
volume, the inclusion of a small quark mass, which explicitly
breaks the chiral symmetry, is essential.

We employ an RG method in order to cover
the relevant range of length scales and pion masses.
Thus, we do not have to rely on either the box size or the pion mass
as an expansion parameter: the RG flow equations remain valid as long as
the model used as input does. The great advantage
is precisely that the RG flow equations 
describe the connection between different momentum scales, and in the
present case also the dependence on the additional scale $1/L$
introduced by the finite volume.

\section{Renormalization Group Flow Equations}
In the present work, we use the two-flavor
quark-meson model, an $O(4)$-invariant 
linear sigma model where the
meson fields $\phi = (\sigma, \vec{\pi})$  are coupled to two
constituent quarks through an $SU(2) \otimes SU(2)$-invariant
interaction term.  
It is an effective model for dynamical chiral symmetry breaking below
a scale $\Lambda \simeq 1.5 \; \mathrm{GeV}$, where a
description of QCD in terms of hadronic degrees of freedom is valid. 
Although the linear sigma-model by itself is not compatible with the 
low-energy $\pi\pi$-scattering data, due to the presence of quarks
the low energy constants of chiral perturbation theory are
reproduced \cite{Jungnickel:1997yu}.

We consider the model in a four-dimensional Euclidean volume with
compact Euclidean space and time directions. 
The bare effective action at the scale $\Lambda$ is
\be
\Gamma_{\Lambda}[\phi]&=& \int d^4 x
\left\{ \bar{q} \gamma \cdot \partial q + g\bar{q}(m_c +
\sigma+i\vec{\tau}\cdot\vec{\pi} 
\gamma_{5}) q +
\frac{1}{2}(\partial_{\mu}\phi)^{2}+U_\Lambda(\phi)\right\}, 
\ee
where the quark mass term $g m_c$ explicitly breaks the chiral symmetry.
At the scale $\Lambda$, the meson potential can be characterized by the values
of two couplings:
\be
U_\Lambda(\phi) &=& \frac{1}{2}m_{UV}^2 \phi^2 + \frac{1}{4}
\lambda_{UV} (\phi^2)^2.
\ee 
In Gaussian approximation, the one-loop effective action for the
scalar fields is
\be
\Gamma[\phi] &=& \Gamma_\Lambda[\phi] - \Tr \log
\left(\Gamma_F^{(2)}[\phi] \right) 
+ \frac{1}{2} \Tr \log
\left(\Gamma_B^{(2)}[\phi] \right),  
\ee
where $\Gamma_B^{(2)}[\phi]$ and $\Gamma_F^{(2)}[\phi]$ are the
inverse two-point functions for the bosonic and the fermionic fields,
evaluated at the expectation value of the mesonic field,
$\phi$. We consider an approximation where the field $\phi$ is
constant over the entire 
volume and the effective action is reduced to an effective potential,
$\Gamma[\phi] = \int d^4 x \;U(\phi)$. 
To regularize the functional traces, we use a Schwinger proper-time
representation of the logarithms. The dependence on a cutoff scale $k$
is introduced through 
an infrared cutoff function $f_a(\tau k^2)$ of the form 
\be
k \frac{\partial}{\partial k} f_a(\tau k^2) &=& - \frac{2}{a!}(\tau
k^2)^{a+1} \exp(-\tau k^2)
\ee  
which satisfies all required regularization conditions
\cite{Schaefer:1999em, Papp:1999he}. 
For reasons of convergence of the momentum integrals, we use $a=2$. 
Such a cutoff function makes it possible to systematically integrate
out only those quantum 
fluctuations around the expectation value which have momenta above the
scale $k$.

We now wish to obtain a renormalization group flow equation for the effective
action, which describes how the couplings in the action change with a
change of the renormalization scale $k$. We arrive at such an equation
by taking the derivative of the regularized expression  
for the effective action with respect to this scale $k$, and then
replacing the bare two-point function from the original
expression by the renormalized two-point functions, which contain the
scale-dependent couplings:
\be
k \frac{\partial}{\partial k} \Gamma_k[\phi] &=& \frac{1}{2} \Tr
\int_0^\infty \frac{d\tau}{\tau} \left[k \frac{\partial}{\partial k}
  f_a(\tau k^2) \right] \exp[- \tau \Gamma^{(2)}_{B, k}[\phi]] \el
&&
-  \Tr \int_0^\infty \frac{d\tau}{\tau} \left[k \frac{\partial}{\partial k}
  f_a(\tau k^2) \right] \exp[- \tau \Gamma^{(2)}_{F, k}[\phi]].
\ee
Note that boundary conditions for the fields are implicit in the
traces. In Euclidean time direction, the sum over bosonic and fermionic
Matsubara frequencies can be done analytically  \cite{Braun:2003ii}
before the zero temperature limit is taken. In infinite volume, the
momentum integrations for the 
spatial directions can be performed as well, and the flow equation becomes
\be
k\frac{\partial}{\partial
  k} U_{k}(\sigma,\vec{\pi}^2, L \to \infty) &=& \frac{k^6}{32 \pi^2}
\Big(-\frac{4  N_{c} N_{f}}{k^2+ M_{q}^{2}} +\sum_{i=1}^{4} 
\frac{1}{k^2 + M_{i}^{2}} \Big). 
\label{eq:flowinfinite}
\ee
In finite volume, the momentum integrations become sums over momentum
modes. Choosing anti-periodic boundary conditions for the  fermionic
fields also in the spatial directions, we define 
\be
p_{F}^{2}=\sum_{i=1}^{d-1}p_{i}^{2} = \frac{4\pi^{2}}{L^{2}}
\sum_{i=1}^{d-1}\left( n_{i}+\frac{1}{2}\right)^{2}, &\;\;& 
p_{B}^{2}=\sum_{i=1}^{d-1}p_{i}^{2} = \frac{4\pi^{2}}{L^{2}}
\sum_{i=1}^{d-1} n_{i}^{2}.
\ee
The RG flow equation for the meson potential in finite volume
then reads
\be
k\frac{\partial}{\partial
  k} U_{k}(\sigma,\vec{\pi}^2,
L)\!\!\!&=&\!\!\!\frac{3}{16}\frac{k^{6}}{L^{3}}\sum_{\vec{n}}\Big(\!-\frac{4
  N_{c} N_{f}}{(k^2+p_F^2+M_{q}^2)^{5/2}} +\sum_{i=1}^{4}
\frac{1}{(k^2 + p_B^2+ M_{i}^2)^{5/2}} \Big)\!. \label{eq:flowfinite}
\ee
The meson masses $M_i^2(\sigma, \vec{\pi}^2)$ are the eigenvalues of
$[U_k^{\prime\prime}(\sigma, \vec{\pi}^2)]^{ij}$,  
the second derivative matrix of the meson potential. In the presence of a
finite current quark mass $g m_c$, the quark mass $M_q^2(\sigma, \vec{\pi}^2) =
g^2[(\sigma_0 + m_c)^2 + 2 m_c (\sigma - \sigma_0) + 
  (\sigma^2 +\vec{\pi}^2 -\sigma_0^2)]$ contains explicitly
symmetry breaking terms. Since symmetry breaking terms appear only
here and are all of this form, we make for the meson potential the ansatz 
\be
U_k(\sigma, \vec{\pi}^2)&=& \sum_{i=0}^{4}
\sum_{j=0}^{[\frac{1}{2}(4-i)]} a_{ij}(k)
(\sigma-\sigma_0(k))^i(\sigma^2 +\vec{\pi}^2 - \sigma_0(k)^2)^j. 
\label{eq:potansatz}
\ee
It depends only on $\vec{\pi}^2$ since the symmetry of the pion
subspace remains unbroken. 

\section{Results of the Numerical Calculations}

We solve the RG flow equations numerically for infinite and finite volume.
In order to do this, we insert 
the ansatz \eqref{eq:potansatz} into the flow equations
\eqref{eq:flowinfinite},\eqref{eq:flowfinite} and obtain a set of coupled,
ordinary differential equations for the couplings 
$a_{ij}(k)$ and the minimum $\sigma_0(k)$. We choose initial values
for the couplings at the scale $k=\Lambda$ and evolve the couplings
until at $k \to 0$ all quantum fluctuations have been integrated
out. Pion mass and pion decay constant are extracted from the
resulting effective potential. 
For the case of finite volume, the sums over the momentum modes need to 
be truncated at a finite number of modes. For the results presented
here, where we use $n_{\mathrm{max}}=40$, the effects of this
truncation are negligible \cite{Braun:2004yk}.

\begin{table}
\begin{tabular}{rrrrrr}
\hline
\tablehead{1}{c}{b}{$\Lambda$\\ $\mathrm{[MeV]}$} &
  \tablehead{1}{c}{b}{$m_{UV}$\\$\mathrm{[MeV]}$} & 
\tablehead{1}{c}{b}{$\lambda_{UV}$\\} & 
\tablehead{1}{c}{b}{$g   m_c$\\$\mathrm{[MeV]}$}  &  
\tablehead{1}{c}{b}{$f_\pi$\\  $\mathrm{[MeV]}$} & 
\tablehead{1}{c}{b}{$m_\pi$\\$\mathrm{[MeV]}$}  \\
\hline
1500 & 779.0 & 60 &  2.10 & 90.38 & 100.8 \\
1500 & 747.7 & 60 &  9.85 & 96.91 & 200.1 \\
1500 & 698.0 & 60 & 25.70 & 105.30 & 300.2 \\
\hline
\end{tabular}
\caption{\label{tab:start} Values for the parameters 
  at the $UV$-scale $\Lambda$ which were used in the numerical
  evaluation. The parameters 
  are obtained by fitting the results of the RG evolution to a
  particular pion mass and the  
  corresponding pion decay constant in infinite
  volume. The physical current quark mass corresponds to $g m_c$.} 
\end{table}
In TAB.~\ref{tab:start}, we give an overview over the three parameter sets
we have used to obtain results for pion masses of $100$, $200$ and
$300\;\mathrm{MeV}$. These parameters are obtained by fitting so that the
results of the RG evolution match a
particular value of the pion mass 
$m_\pi(\infty)$ and the corresponding value of the
pion decay constant $f_\pi(\infty)$ from chPT in infinite volume. With
these fixed 
parameter values, we solve the finite volume RG equations, which
gives a prediction for the volume dependence of $m_\pi(L)$ and
$f_\pi(L)$. The coupling $g$ does not evolve in our approximation and
was set to $g=3.26$, which leads to a reasonable constituent quark mass
for physical values of $f_\pi$ and $m_\pi$.
\begin{figure}[ht!]
\includegraphics[scale=0.80, clip=true, angle=0,
  draft=false]{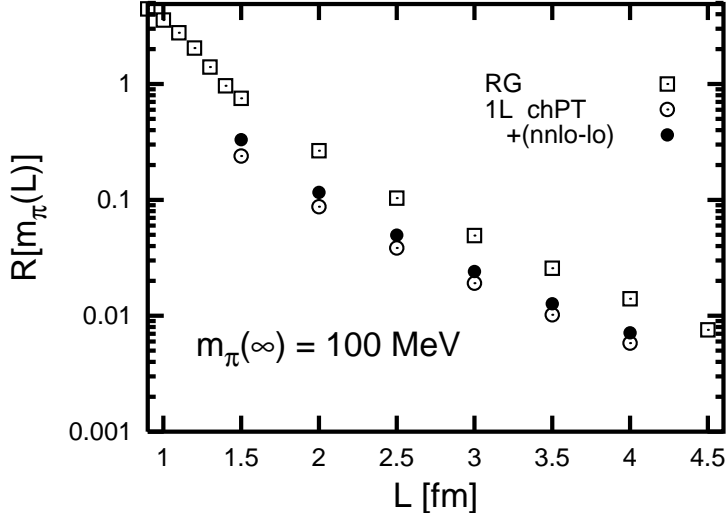} 
\caption{\label{fig:RmpiM_100} Volume dependence of the pion mass for
  a value $m_\pi(\infty) = 100\;
  \mathrm{MeV}$ in infinite volume. We plot the 
  relative shift of the pion mass from its infinite volume limit
  $R[m_\pi(L)] = (m_\pi(L)-m_\pi(\infty))/m_\pi(\infty)$ as a function
  of the size of the volume $L$. For comparison to the RG result, we also
  plot the results from chPT calculations taken from
  \cite{Colangelo:2003hf} for the full one-loop result (1L chPT) and
  with {\it nnlo}-corrections (1L chPT + ({\it nnlo}-{\it lo})).}
\end{figure}
\begin{figure}[ht!]
\includegraphics[scale=0.80, clip=true, angle=0,
  draft=false]{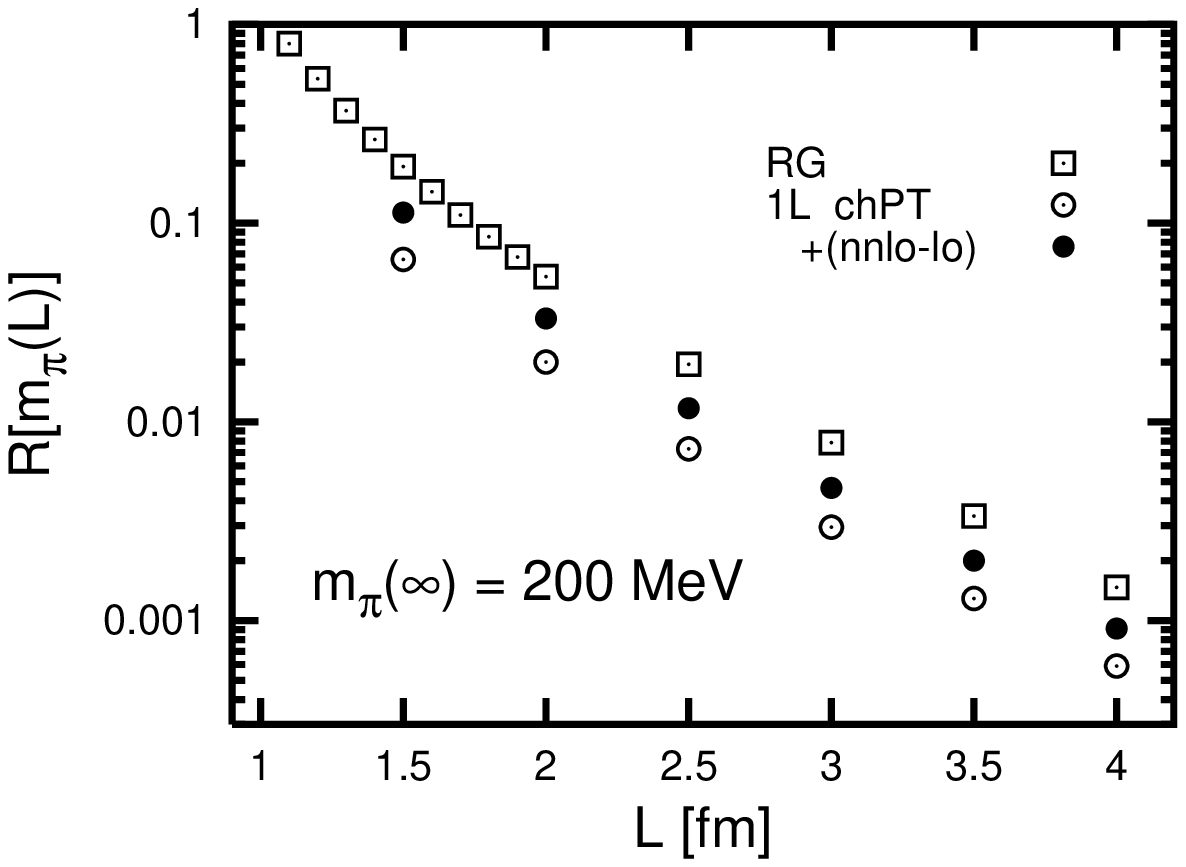}\\ 
\caption{\label{fig:RmpiM_200} Volume dependence of the pion mass for
  a value $m_\pi(\infty) = 200\; \mathrm{MeV}$ in infinite volume. For
  a detailed explanation, see caption of 
  FIG.~\ref{fig:RmpiM_100}. Note the different scales on the axes for
  different values of $m_\pi(\infty)$.} 
\end{figure}
\begin{figure}[ht!]
\includegraphics[scale=0.80, clip=true, angle=0,
  draft=false]{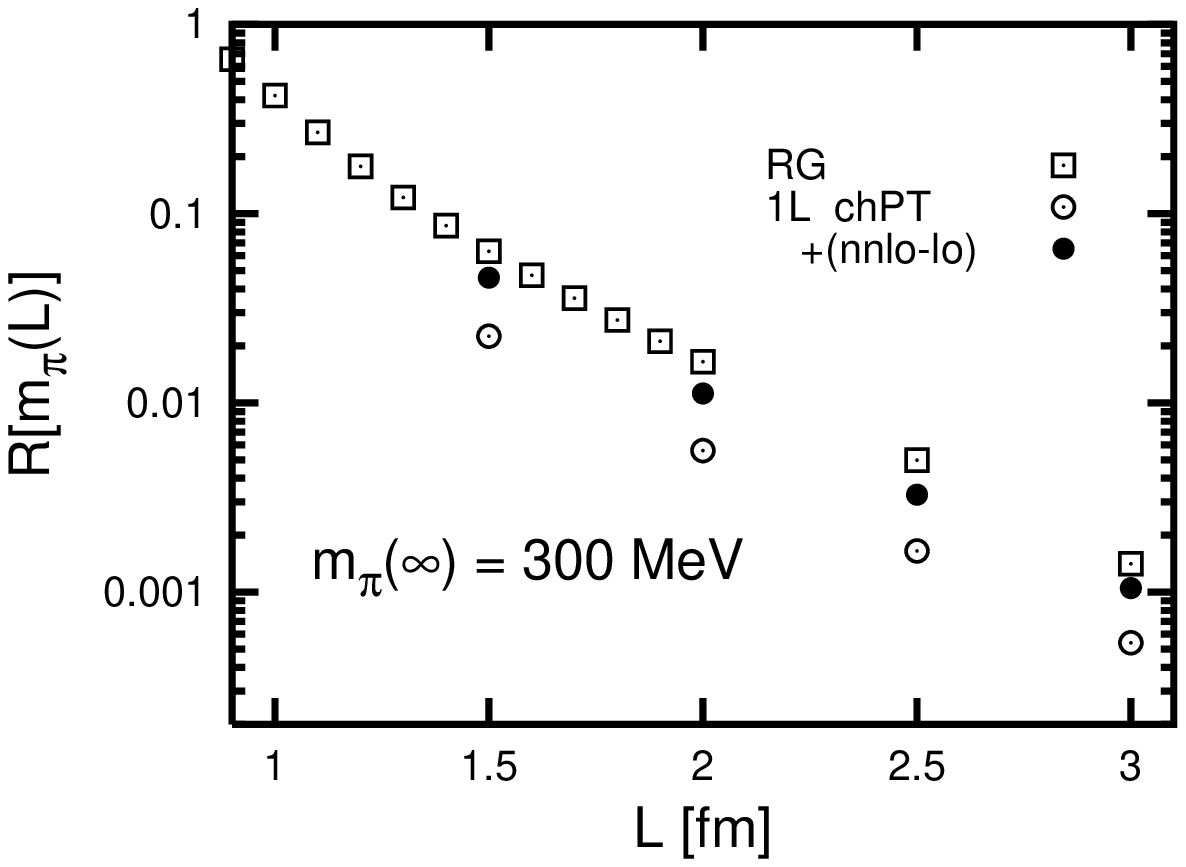}\\ 
\caption{\label{fig:RmpiM_300} Volume dependence of the pion mass for
  a value $m_\pi(\infty) = 300\; \mathrm{MeV}$ in infinite volume. For
  a detailed explanation, see caption of 
  FIG.~\ref{fig:RmpiM_100}. Note the different scales on the axes  for
  different values of $m_\pi(\infty)$.}
\end{figure}
\begin{table}
\begin{tabular}{lllrrrrr}
\hline
\tablehead{1}{c}{b}{$L$} & 
\tablehead{1}{c}{b}{$m_\pi(\infty)$} & 
\tablehead{1}{c}{b}{$m_\pi L$} & 
\tablehead{3}{c}{b}{${R[m_\pi(L)]}$}& 
\tablehead{1}{c}{b}{$\Delta R$}
\\ 
\tablehead{1}{c}{b}{$\mathrm{[fm]}$} &
\tablehead{1}{c}{b}{$\mathrm{[MeV]}$} & &
 \tablehead{1}{c}{b}{{\textmd{RG}}} & 
\tablehead{1}{c}{b}{\textmd{ 1L chPT}} & \tablehead{1}{c}{b}{ +({\it
  nnlo}-{\it lo})} &  \\ 
\hline
2.0 & 100 & 1.01293 & $26.6 \times 10^{-2}$ & $8.74 \times 10^{-2}$ &
$11.6 \times 10^{-2}$ &  $15.0 \times 10^{-2}$\\  
  & 200 & 2.02586 & $5.38 \times 10^{-2}$  & $2.00 \times 10^{-2}$ &
$3.31 \times 10^{-2}$ & $2.07 \times 10^{-2}$ \\ 
  & 300 & 3.03879 & $1.70 \times 10^{-2}$ & $0.56 \times 10^{-2}$&
$1.12 \times 10^{-2}$ & $0.58 \times 10^{-2}$ \\ 
\hline
2.5 & 100 & 1.26616 & $10.37 \times 10^{-2} $ & $3.85 \times 10^{-2}$
& $4.97 \times10^{-2}$ & $5.40 \times 10^{-2}$ \\ 
    & 200 & 2.53233 & $1.95 \times 10^{-2}$ & $0.73 \times 10^{-2}$ &
$1.17 \times 10^{-2}$ & $0.78 \times 10^{-2}$  \\ 
    & 300 & 3.79849 & $5.31 \times 10^{-3}$ & $1.65 \times 10^{-3}$ &
$3.27 \times 10^{-3}$ & $2.04 \times 10^{-3}$  \\  
\hline  
3.0 & 100 & 1.5194 & $4.94 \times 10^{-2}$ & $1.91\times 10^{-2}$ &
$2.41\times 10^{-2}$ & $2.53 \times 10^{-3}$  \\ 
    & 200 & 3.03879 & $7.85 \times 10^{-3}$ & $2.95 \times 10^{-3}$ &
$4.65 \times 10^{-3}$ & $3.20 \times 10^{-3}$ \\  
    & 300 & 4.55819 & $1.76 \times 10^{-3}$ & $0.54 \times 10^{-3}$ &
$1.05 \times 10^{-3}$ & $0.71 \times 10^{-3}$  \\ 
\hline
\end{tabular}
\caption{\label{tab:Rmpi}Values for $R[m_\pi(L)]$, the relative shift 
  of the pion mass in finite volume compared to the value in infinite
  volume (cf. eq.~\eqref{eq:Luscher}). We show results for volume
  sizes
  $L=2.0,\;2.5,\;3.0\;\mathrm{fm}$, and for 
  pion masses $m_\pi(\infty)=100, 
  \; 200, \; 300\;\mathrm{MeV}$. The RG results are compared 
  to the exact one-loop chPT results of
  \cite{Gasser:1987ah} (1L chPT), and to
  the exact one-loop calculation with corrections in
  three-loop order obtained with chPT and L\"uscher's formula
  \cite{Colangelo:2003hf} (1L chPT + ({\it nnlo}-{\it lo})). In the last
  column, the difference $\Delta R$ between the RG result and the
  three-loop corrected chPT result is given.} 
\end{table}
We present our results for the volume dependence of the pion mass in
FIGS.~\ref{fig:RmpiM_100}-\ref{fig:RmpiM_300}
and numerical values for  select volumes in TAB.~\ref{tab:Rmpi}. We
will focus on these 
results for the pion mass, since they are more easily accessible 
on the lattice 
than the pion decay constant. For the results on
$f_\pi(L)$, we refer to \cite{Braun:2004yk}.

FIGS.~\ref{fig:RmpiM_100}-\ref{fig:RmpiM_300} show the relative change
$R[m_\pi(L)]=\frac{m_\pi(L)-m_\pi(\infty)}{m_\pi(\infty)}$ of the pion
mass in finite  
volume $m_\pi(L)$ compared to the value $m_\pi(\infty)$ in infinite
volume as a  function of the volume size $L$. We plot the results for
the pion masses $m_\pi(\infty)=100,\; 200, \; 300\;\mathrm{MeV}$
on a logarithmic scale. For
comparison, we plot in addition the one-loop chPT
result of  \cite{Gasser:1987ah}, and the one-loop chPT result with
corrections to three-loop order obtained with the L\"uscher formula
\cite{Colangelo:2003hf}. 

The relative change of the pion mass decreases with increasing volume
size. The RG results are consistently above the results from chPT.
Higher loop order corrections in chPT increase the mass shift
predicted by chPT, and the difference to the RG result becomes smaller. 
We note that the relative size of the higher loop order corrections to the
one-loop chPT result become larger with larger pion mass. This is not
unexpected, since L\"uscher's formula becomes an increasingly better
approximation with increasing pion mass for a given volume size.
For larger volumes, the differences between the RG result and the
one-loop chPT calculation with the corrections obtained with
L\"uscher's formula drop exponentially as $\exp(- c\, m_\pi L)$ with
$c >0$, compatible with the error estimate from L\"uscher's result. 
As expected, for large volume the slopes of the chPT results and
the RG results in the logarithmic plot are the same, since the mass
shift is then due entirely to pion effects and thus should drop as
$\exp(-m_\pi L)$. In contrast to chPT, the RG calculation can also be
extended to very small volumes, where the chiral expansion is
considered unreliable, and describe the transition into the regime
where chiral symmetry is effectively restored.  

In our model, two effects are responsible for the finite volume
mass shift. Chiral symmetry is broken dynamically and light Goldstone
modes appear when the quark fields form a condensate. The presence of
the additional scale $1/L$ affects this condensation.
In a finite volume, the lowest possible momentum mode of the fermions is
$\sqrt{3}\pi/L$. For small volumes, this acts as an infrared cutoff,
which effectively ``freezes'' the quark fields so that fewer modes can
contribute to the condensate.
The mesonic fields, on the other hand, are much less
affected by the finite volume, since the scale $2 \pi/L$ imposes only
a minimum for the smallest non-zero momentum mode. Fluctuations due to
these light mesonic modes lead to a further decrease of the
condensate, which in turn leads to the observed increase in the masses
of the light mesons. These pion effects dominate for larger volumes,
which explains the convergence of the RG and chPT results for large $L$.

\section{Conclusions}

We have presented a study of finite volume effects on the pion mass
and on the pion decay constant. We have applied renormalization group
methods to a phenomenological model of low-energy QCD with mesons and
constituent quarks as degrees of freedom. We solved the
resulting RG flow equations in finite volume. We stress the importance
of taking explicit symmetry breaking into account and of considering
pions and the sigma as individual degrees of freedom.

We compared our results for the volume dependence of the pion mass to
recent results obtained in 
chiral perturbation theory. While they are compatible for small pion
masses and large volumes, we find from the RG generally a larger mass
shift than the one predicted by chiral perturbation theory. We explain
this for small volumes by the effect of the additional IR cutoff
$L$ on the quark condensation, while for larger volumes the fluctuations
due to light pions are the dominating effect. This view is supported
by the convergence between the RG and chPT results for large volumes. 
Thus, the RG approach in a model including quarks provides a mechanism to
qualitatively understand the volume dependence. The RG results remain
valid even to very small volumes, where the chiral expansion becomes
unreliable. 

An important point that still requires careful investigation is the
influence of the choice of the boundary conditions for the fermionic
fields in the spatial directions. As on the lattice, where
anti-periodic boundary conditions for quarks lead to a larger finite
volume shift than periodic boundary conditions \cite{Aoki:1993gi}, we
observe a dependence of our results on this choice. This will be the
topic of future work.


\begin{theacknowledgments}
We would like to thank the organizers for a very pleasurable
and inspiring meeting, for the hospitality in Belgium and for a thoroughly
enjoyable experience. J.B. would like to thank the GSI for financial support.
\end{theacknowledgments}

\end{document}